\documentclass[usenatbib,twocolumn,useAMS]{mn2e}
\usepackage{graphicx}
\usepackage{latexsym}
\usepackage{dcolumn}
\usepackage{bm}
\input{epsf}
\newcommand{\be}{\begin{equation}}
\newcommand{\ee}{\end{equation}}
\newcommand{\ba}{\begin{eqnarray}}
\newcommand{\ea}{\end{eqnarray}}

\newcommand{\bi}{\begin{itemize}}
\newcommand{\ei}{\end{itemize}}
\newcommand{\bfi}{\begin{figure}
\epsfxsize=9cm
\epsffile}
\newcommand{\efi}{\end{figure}}

\newcommand{\mnras}{MNRAS}
\newcommand{\apj}{ApJ}

\newcommand{\aap}{AAP}
\newcommand{\prd}{PRD}
\newcommand{\prl}{PRL}

\newcommand{\araa}{Annual Review of Astronomy \& Astrophysics}

\newcommand{\aas}{Annals of Applied Statistics}

\title[The optimal weighting function for cosmic magnification]{The optimal weighting function for cosmic magnification
measurement through foreground
galaxy-background galaxy (quasar) cross correlation}
\author[Xiaofeng Yang \& Pengjie Zhang]
{Xiaofeng Yang$^{1,2}$\thanks{E-mail:xfyang@shao.ac.cn},
  Pengjie Zhang$^1$\\
$1$Key Laboratory for Research in Galaxies and
Cosmology, Shanghai Astronomical Observatory, Chinese Academy of
Sciences, \\Nandan Road 80, Shanghai, 200030, China\\
$2$Graduate University of Chinese Academy of Sciences, 19A,
Yuquan Road, Beijing, 100049, China}

\begin{document}
\maketitle

\begin{abstract}
Cosmic magnification has been detected through cross correlation between
foreground and background populations (galaxies or quasars). It has been shown
that weighing each background object  by its $\alpha-1$ can significantly
improve the cosmic magnification measurement
\citep{Menard02,Scranton05}.  Here, $\alpha$ is the logarithmic slope of the
luminosity function of background populations.  However, we find that this
weighting function is optimal only for sparse background populations in which intrinsic clustering is negligible with respect to shot noise.
We derive the optimal weighting function for general case including scale independent and scale dependent weights. The optimal weighting
function improves the S/N (signal to noise ratio) by $\sim 20\%$ for a BigBOSS-like survey and the improvement can reach a factor of $\sim 2$ for  surveys with much denser
background  populations.
\end{abstract}


\section{Introduction}
Weak gravitational lensing directly probes the matter distribution
of the universe (e.g.\citep{Refregier03}) and is emerging as one of the most
powerful probes of dark matter, dark energy \citep{DETF} and the nature of
gravity \citep{Jain08}. By far the most
sophisticated way to measure weak lensing is through cosmic shear, which is
the lensing induced coherent distortion in galaxy shape (\citealt{Fu08} and
references therein). Coordinated  projects on
precision weak lensing
reconstruction through galaxy shapes have been carried out extensively (STEP,
\citealt{Heymans06}; STEP2,
\citealt{Massey07}; GREAT08, \citealt{Bridle09a}; GREAT10, \citealt{Kitching10}).

Alternatively, one can reconstruct weak lensing through
cosmic magnification, namely the lensing induced coherent
distortion in galaxy number density (e.g. \citet{Gunn67,Menard02,Jain03} and
references  therein).
Neglecting high order corrections, the lensed galaxy (quasar)
number overdensity $\delta^{\rm L}_{\rm g}$ is related to the intrinsic overdensity
$\delta_{\rm g}$ by
\be
\label{eqn:number counts}
\delta^{\rm L}_{\rm g}=\delta_{\rm g}+2(\alpha-1)\kappa.
\ee
Here, $\kappa$ is the lensing convergence and $
\alpha(m,z)=2.5~{\rm dlog}\ n(m,z)/{\rm d} m-1 $
is a function of the galaxy apparent magnitude $m$ and redshift
$z$. The number count of galaxy brighter than $m$ is $N(m)=\int^mn(m)dm$. Throughout the paper we use the superscript ``L" to denote the lensed quantity.

Since cosmic magnification does not involve galaxy shape, weak lensing reconstruction through cosmic magnification automatically avoids all problems associated with
galaxy shape. The key step in such reconstruction is to eliminate
$\delta_{\rm g}$, which is often orders of magnitude larger
than the lensing signal $\kappa$. Usually this is done by cross correlating
foreground population (galaxies) and background population (galaxies or quasars) with
no overlapping in redshift \citep{Scranton05,Hildebrandt09,Wang11}.
The resulting cross correlation is
\ba \label{eqn:correlation} \langle
\delta^{\rm L}_{\rm g,f}(\hat{\theta}) \delta^{\rm L}_{\rm
  g,b}(\hat{\theta'})\rangle\approx2(\alpha_{\rm b}-1)\langle \delta_{\rm
  g,f}(\hat{\theta})\kappa_{\rm b}(\hat{\theta'})\rangle\ .
\ea
Throughout the paper we use the subscript ``b" for quantities
of background population and ``f" for that of foreground
galaxies. The above relation neglects a term proportional
to $\langle \kappa_{\rm f} \kappa_{\rm b}\rangle$, which is typically much
smaller than the $\langle \delta_{\rm g,f}\kappa_{\rm b}\rangle$
term.\footnote{This term can be non-negligible or even dominant for sufficiently
high redshift foreground galaxy samples \citep{Zhang06}.}

It is important to weigh the cross correlation measurement appropriately to
improve the S/N (signal to noise ratio).  Since the signal scales as $\alpha-1$,
\citet{Menard02} first suggested to maximize the S/N by weighing each galaxy with
its own $\alpha-1$. This weighting significantly improves the measurement and
a robust $8\sigma$ detection of the cosmic
magnification was achieved for the first time \citep{Scranton05}.

Nevertheless,  we find that, the $\alpha-1$ weighting is
optimal only in the limit where the background galaxy (quasar) intrinsic
clustering is negligible with respect to the shot noise
in background distribution. The statistical errors (noises) are contributed by both shot noise and intrinsic clustering of foreground and background galaxies. In this letter, we derive the exact expression of the weighting function optimal for the cosmic magnification measurement through cross correlation.  The new weighting can further improve
the S/N by $\sim 20\%$ for galaxy-quasar cross correlation measurement in a
BigBOSS-like survey. We can also employ high redshift galaxies instead
  of quasars as  background sources which can have much larger number density
  and smaller  bias. Smaller shot noise results into
  better performance for the derived optimal weighting. The improvement over
  the $\alpha-1$ weighting can reach a factor of $\sim 2$  for
surveys with background population density of $\sim 2/$arcmin$^2$.

Throughout the paper, we adopt the fiducial cosmology as a flat $\Lambda$CDM universe with $\Omega_{\Lambda}=0.728$,
$\Omega_m=1-\Omega_{\Lambda}$, $\sigma_8=0.807$, $h=0.702$ and initial power
index $n=0.961$, which are consistent with WMAP seven years best-fit parameters \citep{Komatsu11}.

\section{THE OPTIMAL WEIGHTING FUNCTION}
We are seeking for an optimal weighting function linearly operating on the
background galaxy (quasar) number overdensity in flux (magnitude) space. Let's
denote  the background galaxy number overdensity of the $i$-th magnitude bin as
$\delta^{(i)}_{\rm g,b}$ and the corresponding weighting function as $W_i$.
\bi
\item The simplest weighting function is scale independent, so the weighted
background galaxy overdensity is
\be
\sum_i W_i\delta^{(i)}_{\rm
  g,b} \ .
\ee
\item We can further add scale dependence in $W_i$ to increase the S/N. The
  new weighting function convolves the density field. For brevity, we
express it in Fourier space as $W_i(\ell)$. The Fourier transformation of the
weighted background overdensity is
\be
\sum_i\sum_\ell W_i(\ell)\tilde{\delta}^{(i)}_{\rm g,b}(\vec{\ell})\ .
\ee
Here, $\tilde{\delta}_{\rm g,b}$ is  the Fourier component of the overdensity
$\delta_{\rm g,b}$. The
weighting function $W_i(\ell)$ is real and only depends on the amplitude of
the wavevector  $\ell\equiv |\vec{\ell}|$. It guarantees  the weighted
overdensity to be real.
\ei

The S/N of the background-foreground galaxy cross-correlation depends on the
weighting function, so we use the subscript ``W'' to denote the S/N after
weighting.  The overall S/N can be conveniently derived in the Fourier space,
\ba\nonumber
\label{eqn:s/n}
\left(\frac{S}{N}\right)_W^2
=&\sum_\ell\left[\frac{\langle C^{\rm
      CM-g}_\ell\rangle_W}{\langle \Delta C^{\rm CM-g}_\ell \rangle_W}\right]^2 \\ \nonumber
=&\sum_\ell\frac{(2\ell+1)\Delta \ell f_{\rm sky}\langle C^{\rm CM-g}_\ell
  \rangle_W^2}{\langle C^{\rm CM-g}_\ell \rangle_W^2+(\langle C_{\rm g,b} \rangle_W+\langle C_{\rm s,b} \rangle_W)(C_{\rm g,f}+C_{\rm
    s,f})} \\
   \label{eqn:s/n0}
=&\sum_\ell\frac{(2\ell+1)\Delta \ell f_{\rm sky}}{1+(C_{\rm g,f}+C_{\rm
    s,f})\frac{\langle b_{\rm g,b} W \rangle^2C_{\rm m,b}+\langle W^2 \rangle C_{\rm
          s,b}}{4\langle W(\alpha_b-1) \rangle^2C_{\kappa_b \rm g}^2}}\ .
\ea
Here, $C^{\rm CM-g}$ is the cosmic magnification-galaxy cross correlation
power spectrum and $\Delta C ^{\rm CM-g}$ is the associated statistical
error. $\langle \cdots\rangle_W$ denotes the weighted average of the
corresponding quantity. We then have
\be
\langle C^{\rm CM-g}\rangle_W=\langle
2(\alpha_{\rm b}-1)W\rangle C_{\kappa_{\rm b}\rm g}\ .
\ee
Here, $C_{\kappa_{\rm b}\rm g}$ is the cross correlation power spectrum between
background lensing convergence and foreground galaxy overdensity. $\langle uv\rangle$ is  the averaged product of $uv$,
\ba
\label{eqn:operator}
\langle uv \rangle\equiv \frac{\sum_i u(m_i)v(m_i)N_{{\rm b},i}}{\sum_i N_{{\rm
    b},i}} \ .
\ea
Here, $N_{{\rm b},i}$ is the number of background galaxies (quasars) in the given
magnitude bin  $m_i-\Delta m_i/2<m<m_i+\Delta m_i/2$.

The S/N scales with  $f^{1/2}_{\rm sky}$ and $f_{\rm sky}$ is the fractional sky
coverage. $C_{\rm s}$ is the shot noise power spectrum and  the weighted one is  $\langle C_{\rm
  s,b}\rangle_W=\langle
W^2\rangle C_{\rm s,b}$. $C_{\rm g}$ is the galaxy power spectrum. We adopt a
simple bias  model for the foreground and background galaxies. We then have
$C_{{\rm g},i}=b_{{\rm g},i}^2C_{\rm m}$ where $b_{{\rm g},i}$ is the bias of
the $i$-th magnitude bin and $C_{\rm m}$ is the corresponding
matter angular power spectrum. The weighted background galaxy power spectrum is
\be
\langle C_{\rm g,b}\rangle_W=\langle
b_{\rm g,b}W\rangle^2C_{\rm m,b}\ .
\ee
\subsection{The scale independent optimal weighting function}
The optimal weighing function $W$ can be obtained by varying the S/N
(Eq. \ref{eqn:s/n}) with respect to $W$ and maximizing it. The derivation is lengthy but
straightforward, so we leave details in the appendix and only present the
final result here.

The optimal weighting function is of the form\footnote{The derived scale
  independent weighting function implicitly assumes no scale dependence in the
galaxy bias. In reality, the galaxy bias is scale dependent and the
application of Eq. \ref{eqn:exact solution10}  is limited. The exact optimal weighting
function applicable to scale dependent bias is given by Eq. \ref{eqn:exact solution11}.}
\ba
\label{eqn:exact solution10}
W=(\alpha_{\rm b}-1)+\varepsilon b_{\rm g,b}\ \ .
\ea
where the scale independent constant $\varepsilon$ is determined by
Eq.~\ref{eqn:exact
  solution0}.  It is a fixed number for the given redshift bin of the given
survey.  In the limit that shot noise of background galaxies overwhelms their
intrinsic clustering ($C_{\rm s,b}\gg C_{\rm g,b}$), $\varepsilon\rightarrow
0$. In this case, the weighting function $\alpha-1$ proposed by
\citet{Menard02} becomes optimal.

\subsection{The scale dependent optimal weighting function}
The weighting function $W$ (Eq. \ref{eqn:exact solution10}) is optimal under the
condition that $W$ is scale independent. If we relax this requirement and
allow for scale dependence in $W$, we are able to maximize the S/N of the
cross power spectrum measurement at each $\ell$ bin.  Clearly, this further
improves the overall S/N.

In this case, $W$ of different $\ell$ bins are independent to each other. This
significantly simplifies the derivation and we are now able to obtain an
analytical expression for $W$,
\ba
\label{eqn:exact solution11}
W(\ell)=(\alpha_{\rm b}-1)+\left[- \frac{\langle (\alpha_{\rm
      b}-1)b_{\rm g,b} \rangle C_{\rm m,b}(\ell)/C_{\rm s,b}}{1+\langle
    b_{\rm g,b}^2\rangle C_{\rm m,b}(\ell)/C_{\rm s,b}} \right]b_{\rm g,b}(\ell).
\ea
This form is similar to Eq.~(\ref{eqn:exact solution10}), except that
it  is now scale dependent. Here again, in the limit $C_{\rm s,b}\gg
C_{\rm m,b}\sim C_{\rm g,b}$, $W\rightarrow \alpha-1$ and we recover the
result of \citet{Menard02}.  This is indeed the case for SDSS background
quasar sample.

\subsection{The applicability}
Are the derived weighting functions (Eq. \ref{eqn:exact solution10} \&
\ref{eqn:exact solution11}) directly applicable to real surveys?
 From Eq. \ref{eqn:exact solution10} \&  \ref{eqn:exact solution11}, it seems
 that we need to figure out $b_{\rm g,b}$ first. Since $b^2_{\rm
  g,b}\equiv C_{\rm g,b}/C_{\rm m,b}$ and  $C_{\rm m,b}$ is not directly
given by the observation, cosmological priors or external measurements
(e.g. weak lensing) are required to obtain the  absolute value of $b_{\rm
  g,b}$. Hence it seems that the applicability of Eq. \ref{eqn:exact solution10} \&
\ref{eqn:exact solution11} is limited by cosmological uncertainty.

However, this is not the case. Eq. \ref{eqn:exact solution11} shows that, it
is the combination $b_{\rm g,b}^2C_{\rm m,b}$ determines $W$. Since $C_{\rm
  g,b}\equiv b_{\rm g,b}^2C_{\rm m,b}$  and $\alpha_b$ are directly observable,
Eq. \ref{eqn:exact solution11} is determined completely by
observations. Closer look shows that Eq. \ref{eqn:exact solution10} is also
determined by the combination $b_{\rm g,b}^2C_{\rm m,b}$, so the corresponding
weighting is determined completely by observations, too. Hence the derived
optimal weighting functions are indeed directly applicable to real surveys.

For ongoing and planned surveys such as CFHTLS,
COSMOS, DES, BigBOSS, LSST, SKA, etc.,  the number density of background
populations (galaxies) can be high and the intrinsic clustering can be non-negligible or
even dominant comparing to shot noise.  In next section we will quantify the
improvement of the optimal weighting functions for a BigBOSS-like survey and
briefly discuss implications to surveys with even denser background
populations.

\section{The improvement}
\begin{center}
\begin{table*}
\caption{Improving the cosmic magnification measurement by the optimal
  weighting function. The target survey is BigBOSS. The terms on the left side
  of the arrows are the estimated S/N using the  weighting $\alpha-1$. The
  ones on the right side are the S/N using the optimal weighting function,
  where the ones on the left side of $``/"$ are what expected using the scale
  independent weighting (Eq.~\ref{eqn:exact
    solution10}), and the ones on the right are what expected using  the scale
  dependent weighting (Eq.~\ref{eqn:exact solution11}). The improvement
  depends on the bias dependence on galaxy luminosity. To demonstrate such
  dependence, we adopt a toy model $b_Q\propto F^\beta$ and investigate
  different values of the parameter $\beta$.  In general, the
  optimal weighting function improves the S/N by $10\%$-$20\%$ for
  BigBOSS, whose background quasar density is $\sim 0.02/$arcmin$^2$. The improvement can reach a factor of $\sim 2$ for surveys with
  background (galaxy) populations reaching surface density of $\sim 2/$arcmin$^2$. }
\label{tab:prediction0}
\begin{tabular}{cccccc}
\hline\hline
Flux dependence of quasar bias model & $\beta=0$ & 0.1 & 0.2 & 0.3 \\\hline
Detection significance of $\rm LRG\times quasar$  &  111.2$\rightarrow$129.0/136.8 & 109.8$\rightarrow$126.3/133.9 & 108.3$\rightarrow$123.7/131.1 & 106.5$\rightarrow$119.3/126.7 \\
Detection significance of $\rm ELG\times quasar$  & 94.3$\rightarrow$106.4/110.3  & 93.2$\rightarrow$104.5/108.6  & 92.0$\rightarrow$102.2/106.7 & 90.5$\rightarrow$99.1/104.1 \\\hline
\end{tabular}
\end{table*}
\end{center}

BigBOSS\footnote{http://bigboss.lbl.gov/}  is a planned
spectroscopic redshift survey of $24000\ {\rm deg}^2$  (BigBOSS-North plus
South). Cosmic magnification can be measured by BigBOSS through LRG (luminous
red galaxy)-quasar and ELG (emission line galaxy)-quasar cross correlations.
In principle, it can also be
measured  through LRG-ELG cross correlation. But the interpretation of the
measured cross correlation signal would be complicated by the intricate selection
function of ELGs \citep{Zhu09}.  In the current paper, we only consider the
LRG-quasar and ELG-quasar cross correlations.

There are some uncertainties in the BigBOSS
galaxy (quasar) redshift evolution, flux distribution and intrinsic clustering.  To
proceed, we will take a number of simplifications.  So the absolute S/N of
cross correlation measurement that we calculate is by no means  accurate. But
our calculation should be sufficiently robust to demonstrate  the relative
improvement of the exact optimal weighting function over the previous one.

 The LRG and ELG luminosity functions are calculated based on
 the BigBOSS white paper \citep{Schlegel09}. The comoving number density of
 LRG and ELG is $3.4\times10^{-4}(h/\rm Mpc)^3$, then we have $1.1\times10^7$
 LRGs
 in the redshift range of $z=0.2-1$ and $3.3\times10^7$ ELGs in the redshift
 range of $z=0.7-1.95$.  Clustering of LRGs evolves slowly, so we
 adopt LRG bias as $b_{\rm g,f}(z)=1.7/D(z)$ \citep{Padmanabhan06}. Here $D(z)$
 is the linear density growth factor and is normalized such that
 $D(z=0)=1$. Existing knowledge on clustering of ELGs is
 rather limited. So we simply follow \citet{Padmanabhan06} and approximate the
 ELG bias as $b_{\rm g,f}=0.8/D(z)$. 

For the luminosity function of background
quasars, we adopt the LDDE (Luminosity dependent density evolution) model with
best fit parameters from \citet{Croom09}. The magnitude limit is $g=23$, then
we have $2.1\times10^6$ quasars in the redshift range of $z=2-3.5$. We choose
a redshift gap ($z_{\rm b,min}-z_{\rm f,max}=0.05$) such that the
intrinsic cross correlation between foreground and background populations can
be safely neglected.  We adopt a bias model for  quasar clustering, with $b_{\rm Q}(z)=0.53+0.289(1+z)^2$
from the analysis of $3\times10^5$quasars \citep{Myers07}.

 The S/N depends on many issues and can vary from $90$-$140$
 (Table~\ref{tab:prediction0}). The S/Ns of LRG-quasar and ELG-quasar  correlations are comparable because of a consequence of several competing factors including the lensing efficiency, galaxy surface density and clustering. Nevertheless, a robust conclusion is that BigBOSS  can measure cosmic magnification through galaxy-quasar
cross correlation measurement with high precision. Given such high S/N and
accurate redshift available in BigBOSS, it is feasible to directly measure the
angular diameter distance from such measurement by the method of
\citet{Jain03b,ZhangJun05,Bernstein06}.

Unambiguous improvement in the cross correlation measurement by
using our optimal weighting (Eq. \ref{eqn:exact solution10} \&
  \ref{eqn:exact solution11}) is confirmed, as shown in
  Table~\ref{tab:prediction0}.   The  S/N is improved by $\sim 15\%$ by using
  the scale independent optimal weighting (Eq. \ref{eqn:exact solution10}) and
  by $\sim 20\%$ by using the scale dependent one (Eq. \ref{eqn:exact solution11}).

We further investigate the dependence of the above improvement on the flux
dependence of quasar bias. We adopt  a toy model with $ b_{\rm Q}(z,F)=b_{\rm
  Q}(z)(F/F^*)^\beta$. Here, $F^*$ is the flux corresponding to that of the $M^*$ in the quasar luminosity function.  $\beta$ is an adjustable parameter and we will try $\beta=0, 0.1, 0.2, 0.3$, then the corresponding parameters in scale independent weighting are $\varepsilon =0.049, 0.048, 0.047$ and $0.045$. Larger value of
$\beta$ ($\geq0.4$) leads to too large quasar bias ($\geq10$) and hence will
not be investigated here.  Table~\ref{tab:prediction0} shows consistent
improvement by our optimal weighting functions. Hence, despite uncertainties
in quasar modeling,  we expect our optimal weighting function to be useful to improve
 cosmic magnification  measurement in the BigBOSS survey.

Nevertheless the improvement is only moderate. The major reason is that, even
for BigBOSS, the quasar sample is still sparse, with a surface number density
$\sim 0.02/$arcmin$^2$.  Hence shot noise dominates over the intrinsic
clustering. Indeed, we find that  typically  $C_{\rm m,Q}/C_{\rm s,Q}\la
0.1$. For imaging surveys like CFHTLS, COSMOS, DES and LSST, we can also use
high redshift
galaxies as background galaxies to correlate with low redshift foreground
galaxies. For these surveys,  high redshift galaxy population (e.g. with
$z>1$-$2$) can reach surface number density $\sim 0.2$-$2/$arcmin$^2$ or even
higher. So the shot noise in these surveys can be suppressed by a factor of
$10$-$100$ or more.  The overall improvement of our optimal weighting would be
larger.

To demonstrate these further improvements,  we hypothetically increase the
surface density of BigBOSS quasars by a factor of 10 and 100 respectively, but
keep $\beta=0$ and all other parameters unchanged. Shot noise will be decreased by a
factor of 10 and 100 correspondingly. The scale independent weighting parameter $\varepsilon$ can reach 0.12 and 0.22 respectively. For the first case, the S/N is improved by  $\sim 38\%$ for the scale independent optimal weighting and by $\sim
51\%$ for the scale dependent one.  For the second case, the improvement
is  $\sim 72\%$ for the scale independent optimal weighting and is $\sim
94\%$ for the scale dependent one. It is now clear that for measuring cosmic
magnification through cross correlation between foreground and background galaxies
of many existing and planned surveys, one should
adopt the optimal weighting function derived in this letter.

\section{Summary}
We have derived the optimal weighting functions for cosmic magnification
measurement through cross correlation between foreground and background
populations, for scale independent and scale
  dependent weights respectively. Our weighting functions outperform the commonly used
weighting function $\alpha-1$ by $\sim20\%$ for a BigBOSS-like survey and by
larger factors for surveys with denser
background populations. Hence we recommend to use our optimal weighting
function for
cosmic magnification measurement in BigBOSS, CFHTLS, COSMOS, DES, Euclid,
LSST, SKA, WFIRST, etc.

\section{Acknowledgment}
This work is supported in part by the one-hundred talents program of the
Chinese Academy of Sciences, the national science foundation of China
(grant No. 10821302, 10973027 \&  11025316), the CAS/SAFEA
International Partnership Program for  Creative Research Teams and the 973
program grant No. 2007CB815401.

\appendix
\section{Deriving the optimal weighting function}
In Section~2, we give the optimal weighting function without derivation. Here,
we present a brief derivation for the scale independent weighting
function. The derivation of the scale dependent weighting function is similar,
but simpler.

Maximizing the S/N requires the variation of $(S/N)^2$ with respect to $W$ to
be zero. From this condition, we
obtain
\ba\nonumber
\label{eqn:derivative12}
W&=\frac{\sum_\ell(2\ell+1)\Delta \ell\eta (\nu \langle b_{\rm
      g,b}W \rangle^2+\langle W^2
\rangle)\big/(1+F)^2}{\langle W(\alpha_{\rm b}-1) \rangle \sum_\ell(2\ell+1)\Delta \ell \eta\big/(1+F)^2}(\alpha_{\rm b}-1)\\
&-\frac{\langle W(\alpha_{\rm b}-1) \rangle \langle b_{\rm g,b}W \rangle
  \sum_\ell(2\ell+1)\Delta \ell\eta\nu\big/(1+F)^2}{\langle W(\alpha_{\rm b}-1)\rangle \sum_\ell(2\ell+1)\Delta \ell\eta\big/(1+F)^2}b_{\rm g,b}.
\ea
Here, for brevity, we have denoted  $\eta=C_{\rm s,b}(C_{\rm g,f}+C_{\rm
    s,f})/C_{\kappa_b
    g}^2$, $\nu=C_{\rm m,b}/C_{\rm s,b}$ and  $F_{(W)}=\eta[\nu\langle b_{\rm g,b} W
  \rangle^2+\langle W^2 \rangle]/[4\langle W(\alpha_{\rm b}-1) \rangle^2]$.

Noticing that the coefficients of $\alpha_{\rm b}-1$ and $b_{\rm g,b}$ only
involve the weighted average of $W$ and taking the advantage that the optimal $W$
remains optimal
after a constant (flux independent) scaling, we are able to seek for the
solution of the form $W= (\alpha_{\rm b}-1)+\varepsilon b_{\rm
  g,b}$, with $\varepsilon$ a constant to be determined.  Plugging it into the
above equation,  we obtain the equation of $\varepsilon$,

\ba
\label{eqn:exact solution0}
\varepsilon=-\frac{\langle (\alpha_{\rm b}-1)b_{\rm
  g,b}\rangle}{\sum_\ell\frac{(2\ell+1)\Delta
    \ell\eta}{(1+F)^2}\bigg/\sum_\ell\frac{(2\ell+1)\Delta \ell\eta\nu}{(1+F)^2}+\langle
  b_{\rm
  g,b}^2\rangle}\ .
\ea
Noticing that $F$ depends on $\varepsilon$. The above equation can be solved
numerically to obtain the solution of $\varepsilon$.

In the case of scale dependent weighting, each $W(\ell)$ are independent to
each other. Through similar derivation, one can show that
\ba
\label{eqn:exact solution5}
W(\ell)=(\alpha_{\rm b}-1)+\left[-\frac{\nu\langle (\alpha_{\rm b}-1)b_{\rm
  g,b}\rangle}{1+\nu\langle b_{\rm
  g,b}^2\rangle}\right]b_{\rm
  g,b}\ .
\ea

\end{document}